\documentclass[aps,prb,preprint,showpacs]{revtex4}

\usepackage{graphicx}
\usepackage{subfigure}
\usepackage{dcolumn}
\usepackage{bm}

\begin{document}

\title{Complex edge effects in zigzag graphene nanoribbons due to hydrogen loading}
\author{Sumanta Bhandary}
\affiliation{Department of Physics and Astronomy, Uppsala University, Box 516,
 751\,20 Uppsala, Sweden}
 \author{Mikhail I. Katsnelson}
\affiliation{Radboud University Nijmegen, Institute for Molecules
and Materials, Heyendaalseweg 135, 6525 AJ Nijmegen, The
Netherlands}
\author{Olle Eriksson}
\affiliation{Department of Physics and Astronomy, Uppsala University, Box 516,
 751\,20 Uppsala, Sweden}
\author{Biplab Sanyal\cite{bs}}
\affiliation{Department of Physics and Astronomy, Uppsala University, Box 516,
 751\,20 Uppsala, Sweden}

\date{\today}

\begin{abstract}
We have performed density functional calculations as well as
employed a tight-binding theory, to study the effect of
passivation of zigzag graphene nanoribbons (ZGNR) by Hydrogen. We show
that each edge C atom bonded with 2 H atoms open up a gap and
destroys magnetism for small widths of the nanoribbon. However, a
re-entrant magnetism accompanied by a metallic electronic
structure is observed from 8 rows and thicker nanoribbons. The
electronic structure and magnetic state are quite complex for this
type of termination, with sp$^3$ bonded edge atoms being
non-magnetic, whereas the nearest neighboring atoms are metallic
and magnetic. We have also evaluated the phase stability of
several thicknesses of ZGNR, and demonstrate that sp$^3$ bonded
edge atoms, with 2 H atoms at the edge, should be stable at
temperatures and pressures which are reachable in a laboratory
environment.
\end{abstract}

\pacs{73.22.Pr, 75.70.Ak, 73.20.Hb, 81.05.U-}


\maketitle
\section{Introduction}
The physical and chemical properties of graphene have lately been
studied intensely.\cite{r1,r2,r3,r4,r5,r6} One of the reasons for
this is the high electron mobility of this newly discovered
material and the potential for scaling electronics devices to true
nano-sizes. Hence this novel material holds exciting promises for
applications in micro-electronics. For such applications, the
possibility to create a band gap is important, and chemical
functionalization have indeed been reported where the electronic
properties were modified.\cite{coleman,sanyal,jafri,sofo,elias,seb}
Another avenue towards realizing this is through geometrical
confinement, such as that provided by graphene nano-ribbons.

Nanoribbons of graphene, especially the zigzag edge terminated
ones, have recently been under focus, both from experimental and
theoretical studies. For instance, the possibility of breaking
spin-degeneracy\cite{son} was shown, at least as given by first
principles theoretical calculations. In this work an
anti-ferromagnetic state across the nano-ribbon was found to have
the lowest energy, with a resulting small gap opening up at the
Fermi level. Furthermore, the mobility of atoms at the edge was
studied experimentally in Ref.~\onlinecite{jia}, where an
edge-reconstruction was shown to be driven by Joule heating. Also,
the dynamics of atoms at the edges of carbon nano-ribbons was
investigated in Ref.~\onlinecite{girit} where, in addition,
theoretical modeling based on kinetic Monte Carlo simulations were
presented. These studies show that the coupling between geometry
and electronic structure, and even magnetism, is strong in carbon
nano-ribbons. It is important to note that recently the edge
structure of zigzag graphene nanoribbon (ZGNR) was questioned, and
instead theoretical and experimental evidence for a reconstructed
edge geometry were reported, with alternating pentagons and
heptagons at the edge, in a geometry called the reczag
structure.\cite{koskinen1,koskinen2}

In graphene the C atoms are sp$^2$ bonded. Hence, an edge atom
would have dangling bonds, which under any realistic circumstances
would be saturated by an atom or molecular species. In order to
make the theoretical modeling more realistic, a termination of the
edge C atoms with one or two hydrogen atoms per carbon atom has
been considered.\cite{srolovitz,wassmann} In both these works the
Gibbs energy was calculated as a function of, among other things,
the temperature and pressure of the reference state of the H
atoms, i.e. an H$_2$ gas. The edge termination of one H atom per C
atom continues the sp$^2$ bonding of the graphene bulk, with a
resulting planar geometry. This is normally referred to as 1H
termination. The alternative termination, with two H atoms per
edge carbon atom provides a tetrahedral geometry of the C edge
atoms, with sp$^3$-bonding for the edge atoms. This termination is
referred to as 2H termination and the stability of the 1H and 2H
terminations can be calculated as a function of pressure and
temperature of the reference state of the H atoms.

The magnetic properties of 1H terminated graphene have been
studied theoretically to a large extent, and it is found that a
spin-polarized solution gives a lower total energy compared to a
spin-degenerate calculation. The magnetic properties of 2H
terminated ZGNR have also been studied, and are found to be quite
exotic, with the very outermost C atoms being spin-degenerate and
having a gap of the electronic structure. The nearest neighboring
C atoms have however a spin-polarized solution, as shown by
tight-binding theory\cite{kusakabe,lee} and first principles
calculations of an 8 layer ZGNR.\cite{XuYin} Here we extend these
studies to investigate various thicknesses of the ZGNR, and the
stability, electronic structure and magnetic properties of 1H and
2H terminated ZGNR.

The paper is organized as the following. In section II, we have
discussed the computational details used in this study followed by
the results and discussions in section III . We have shown a model
calculation in the support of our ab initio results in section IV
followed by the conclusions.

\section{Computational details}
First principles density functional calculations have been
performed using the Quantum Espresso code \cite{qe} using plane
wave basis sets and pseudopotentials. The exchange correlation
potential has been approximated by using PBE functional\cite{pbe}
in the generalized gradient approximation (GGA).  We have tested
the convergence with respect to the basis-set cutoff energy and a
value of 80 Ry. has been considered for all results shown in this
paper. A 80x1x1 k-points set has been used for the one dimensional
Brillouin zone. Structural optimizations have been done by
minimizing the Hellmann-Feynman forces on each atom with a
tolerance of 0.001 Ry./a.u. We have relaxed the structures with
basis set cutoff energy value of 60 Ry. and one dimensional
k-point mesh of 60x1x1. These parameters give very similar results
as the ones calculated with 80 Ry. cutoff energy and 80x1x1
k-point mesh. For all cases, spin-polarized calculations were
performed. Different widths (3 to 20 rows) of ZGNRs have been
considered in our calculations. The computational unit cell has
been chosen to be 20 \AA~long in the direction perpendicular to
ZGNR plane and at least 15 \AA~along the y direction while the
nanoribbon is infinite in the x-direction.

\section{Results}
The dangling bonds of edge C atoms of nanoribbons are extremely
reactive and need to be saturated, if one has the ambition to
model a realistic situation. There are several possibilities of
edge termination.  We concentrate on H termination at the edges as
this seems to be one of the most stable configurations owing to
its simple planar structure. Although the structure remains
planar, C atoms rearrange themselves on relaxation and  reach a
bond-distance close to 1.42~\AA~ (which is the bulk graphene bond
length) in the middle of the ribbon. We carried out calculations
for 1H termination with different widths, from 3 rows up to 20
rows. We observe that 1H terminated ZGNR is metallic and
spin-polarized with an average  moment of  ~0.25 $\mu_{B}$/C atom
in agreement with previous DFT calculations\cite{louieprl}. We
note however that to date, there is no firm experimental evidence
of magnetism in GNRs. This could be due to that the edge atoms
undergo a structural distortion, which quenches
magnetism\cite{koskinen1}, or due to the one-dimensional geometry
of the spin-polarized atoms, which according to the Mermin-Wagner
theorem should not order at any finite temperature. At finite
temperatures, instead of ferromagnetism, a superparamagnetism
should appear, with an enhancement factor for the magnetic
susceptibility, roughly, 8 at room temperature, according to the
calculations\cite{yaz_kats}. There is however another scenario,
where the edge C atoms have all their bonds saturated, so that the
absence of the lone pair electron results in a nonmagnetic state.
The relevant situation is when two H atoms terminate a C atom. The
chemical binding for this edge C atom is an sp$^3$ hybrid, with
two C atoms and two H atoms as nearest neighbors. It is this
termination we explore here, both in terms of the energetics, but
also the electronic structure and magnetic properties.

For the termination of edges with 2 H atoms per C, the formation
energies have been defined in two ways:
\begin{eqnarray*}
E_{f}^{(1)}=E(G2H)-[E(G1H)+E(H_{2})]\\
E_{f}^{(2)}=E(G2H)-[E(G1H)+2[E(H)+(B.E./H)_{exp}],
\end{eqnarray*}
where E(G2H) and E(G1H) are the total energies for the zigzag
graphene nanoribbons' edges terminated with 1 H and 2 H atoms per
edge, respectively. $E(H_{2})$ and $E(H)$ are the calculated
energies for a H$_{2}$ molecule and a H atom, respectively.
$(B.E./H)_{exp}$ is the experimental binding energy per H atom of
the H$_2$ molecule, and it amounts to 2.24 eV. Our calculated
value is 2.27 eV, which is in very good agreement with the
experimental one.

\begin{figure}[h]
\includegraphics[scale=0.35]{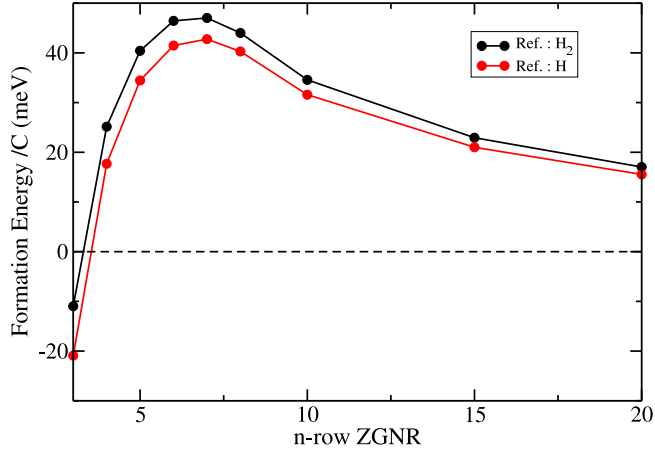}
\caption{(Color online) Calculated formation energies according to
the equations shown in the text for different widths of
nanoribbons. Ref.:H$_{2}$ and Ref.:H indicate the calculated
values using the expressions for $E_{f}^{(1)}$ and $E_{f}^{(2)}$
respectively.} \label{fig1}
\end{figure}

\begin{figure}[h]
\includegraphics[scale=0.35]{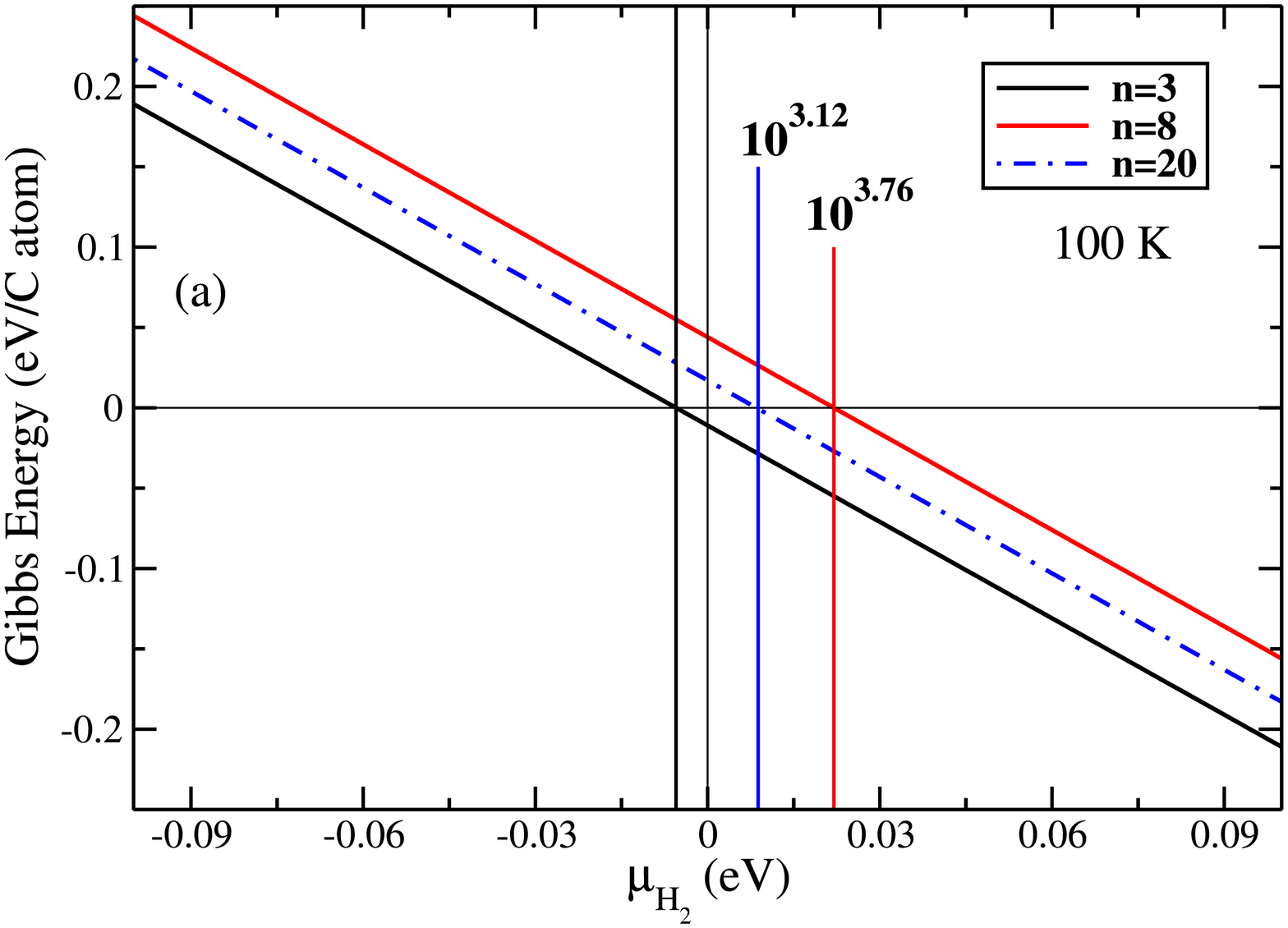}
\includegraphics[scale=0.35]{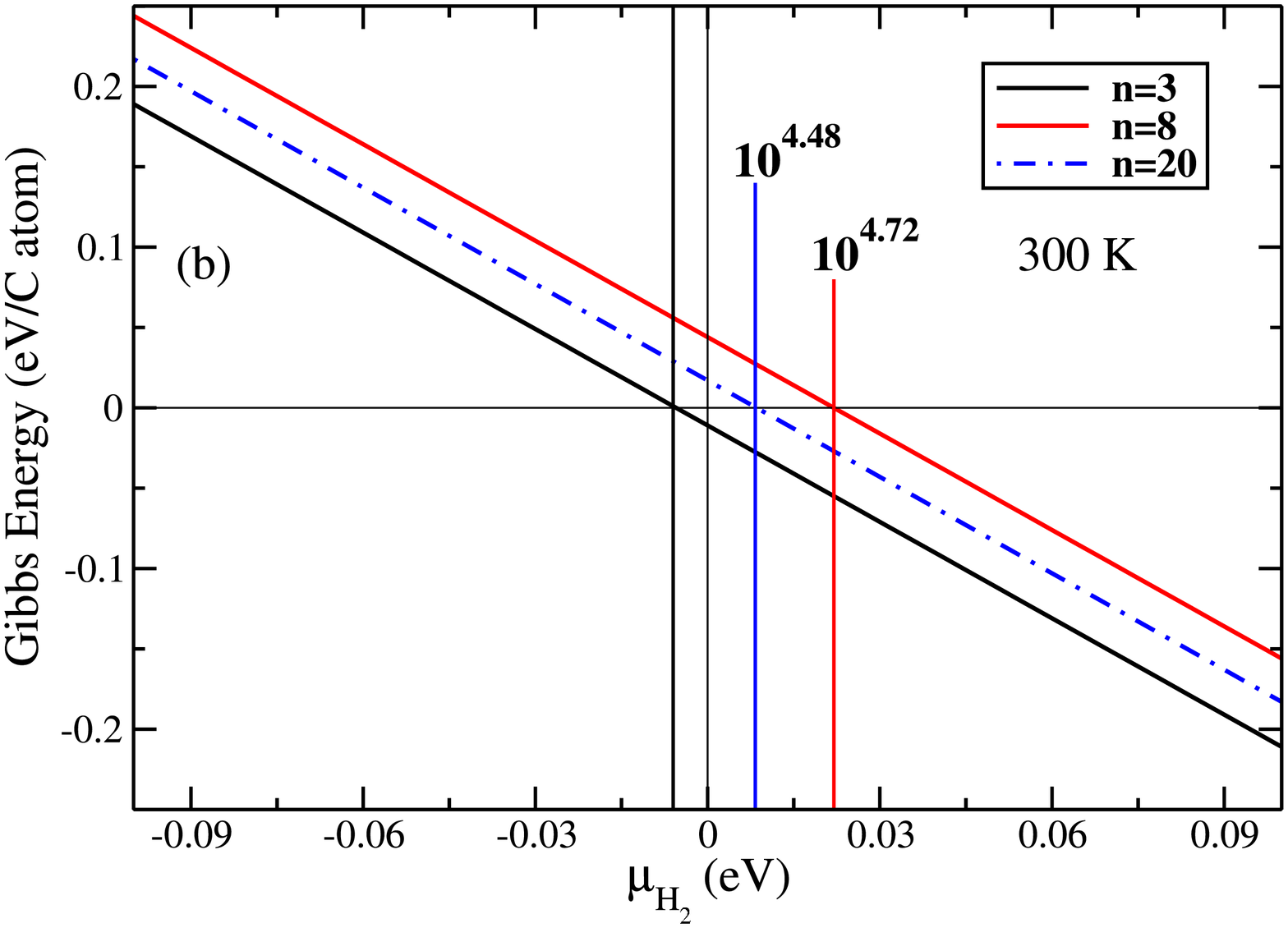}
\caption{(Color online) Calculated Gibbs formation energies for
n=3, 8 and 20 thick ZGNR, at (a) 100 K and (b) 300 K as a function
of chemical potential of hydrogen molecule according to the
equations shown in the text. For the sake of clarity, the vertical
bars are shown to indicate the corresponding chemical potentials
for which the Gibbs free energies become negative. The numbers
indicate the pressures in bar corresponding to the vertical bars.}
\label{fig2}
\end{figure}

The formation energies are shown in Fig.~\ref{fig1} using the two
equations presented above. These energies indicate that the
formation of 2H terminated edges is probable. The formation
energies tend to decrease with the increase in the width of the
nanoribbon whereas for 3-rows ZGNR it is at T=0 K spontaneously
formed. In order to investigate the influence of finite
temperature/elevated gas pressure, we have evaluated the Gibbs
free energy of the reference phase, the gas phase of the H atoms.
Hence, we have calculated the Gibbs free energy as a function of
the chemical potential of the hydrogen molecule, according to the
following formula given by Wassmann {\it et al.}\cite{wassmann}.
\begin{eqnarray*}
G_{H_{2}}=E_{H_{2}}-2 \mu_{H_{2}} \\
E_{H_{2}} = E(G2H)-[E(G1H)+E(H_{2})] \\
\mu_{H_{2}} = H^{0}(T) - H^{0}(0) -TS^{0}(T) + k_{B}T~ln(\frac{P}{P^{0}})
\end{eqnarray*}
In the above equations, $\mu_{H_{2}}$, $H$, $S$, $P$ and $k_{B}$
are the chemical potential, enthalpy, entropy, pressure and
Boltzmann constant respectively. The values for the entropies and
enthalpies are taken from the tabular data presented in Ref.~\onlinecite{chase}.
$P^{0}$ is the reference pressure taken to be 0.1 bar according to
the tabular data. $E(G2H)$, $E(G1H)$ and $E(H_{2})$ have been
defined above. The results are shown in Fig.~\ref{fig2} for 100 K
and 300 K. One can observe the stability of the nanoribbons
terminated with 2H at certain pressures indicated in the plots.
The 20 rows-ZGNR requires lower pressure of the H$_2$ gas,
compared to the 8-rows ZGNR, both for 100 K and 300 K case (the 2H
terminated ZGNR is stable at negative values in the plot). As seen
in the plots, the chemical pressure of molecular hydrogen required
to stabilize a 2H terminated ZGNR is within the range of
laboratory values.

\begin{figure}[h]
\begin{center}
\includegraphics[width=90mm,height=70mm]{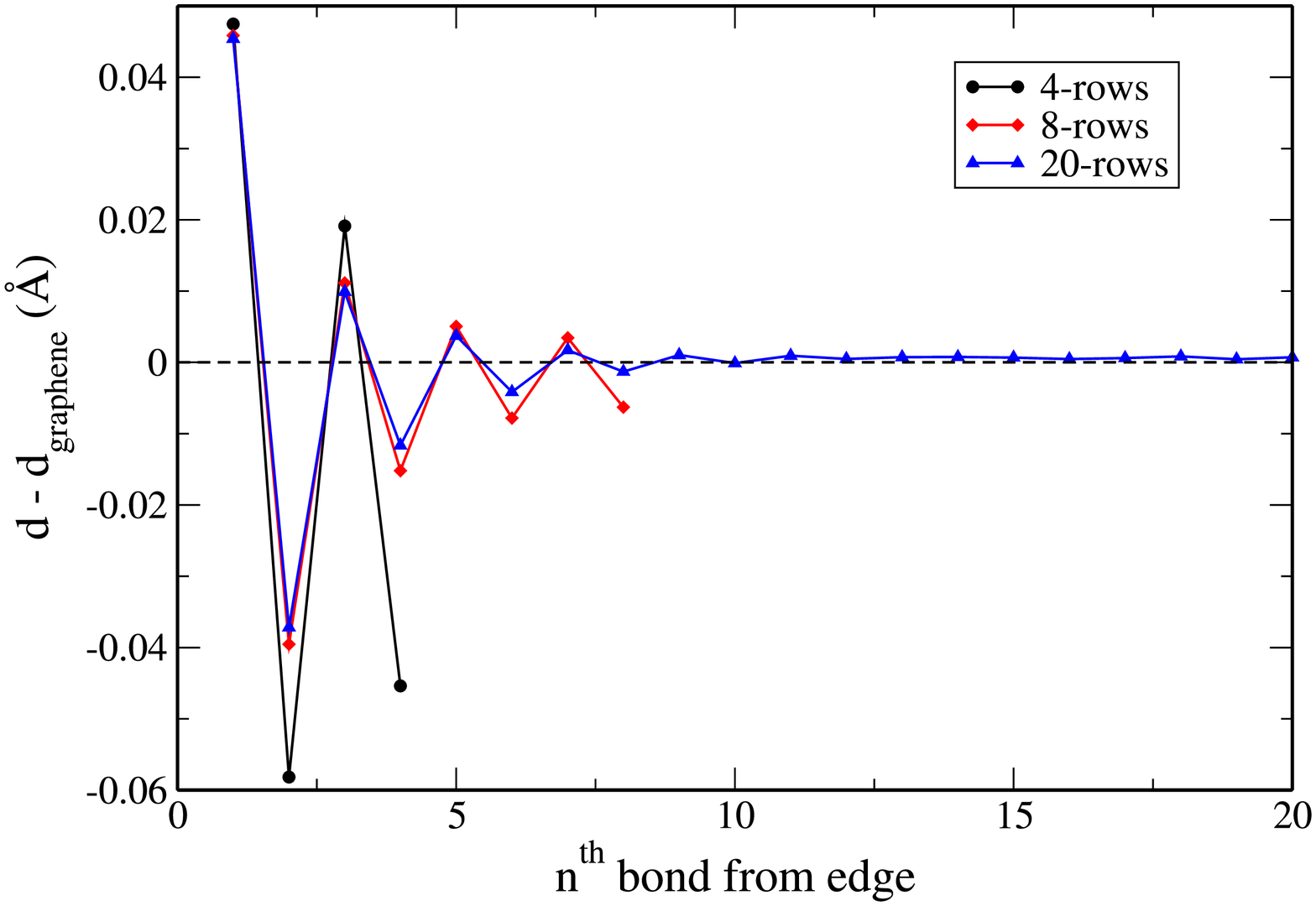}
\caption{(Color online) Calculated C-C bond lengths relative to an
ideal 2D graphene layer (1.42~\AA~) for different bonds from the
edge towards the center of ZGNRs, with thickness 4, 8 and 20
rows.} \label{fig3}
\end{center}
\end{figure}

With two H atoms terminating a C atom, the chemical bonds are now
completely saturated, which results in a vanishing magnetic moment
of this atom. In addition, we observe the appearance of
distortions in the ZGNR plane. The sp$^{2}$ planar structure now
becomes buckled, making a sp$^{3}$ like structure with an angle of
102$^{o}$ at the edge carbon. The C-C bond length increases at the
edge (see Fig.~\ref{fig3}) and this bond-stretching remains almost the same,
for all widths of the ribbon. However, the inner bond-lengths have
an oscillatory behavior as as one traverses towards the center of
the ribbon (Fig.~\ref{fig3}). The C-C bond length of 1.42 \AA~ in sp$^{2}$ bonded
graphene is recovered in the middle of the ribbon. For the 20 rows nanoribbon, 
the bond-length reaches
the ideal  value in bulk graphene around the 15$^{th}$ bond from
the edge. It is obvious that the edge sp$^{3}$ structure has a pronounced effect on
smaller ribbons by perturbing the whole structure where as for
wider ribbons, this perturbation dies out and we get a planar
valley along with sp$^{3}$  distorted edge as seen in
Fig.\ref{fig3}. Besides the edge region with a CH$_{2}$ unit, the
wider ribbons have similar features as in 1H terminated ribbons
which are of planar sp$^{2}$ type throughout their structures.

\begin{figure}[h]
\begin{center}
\includegraphics[width=80mm,height=65mm]{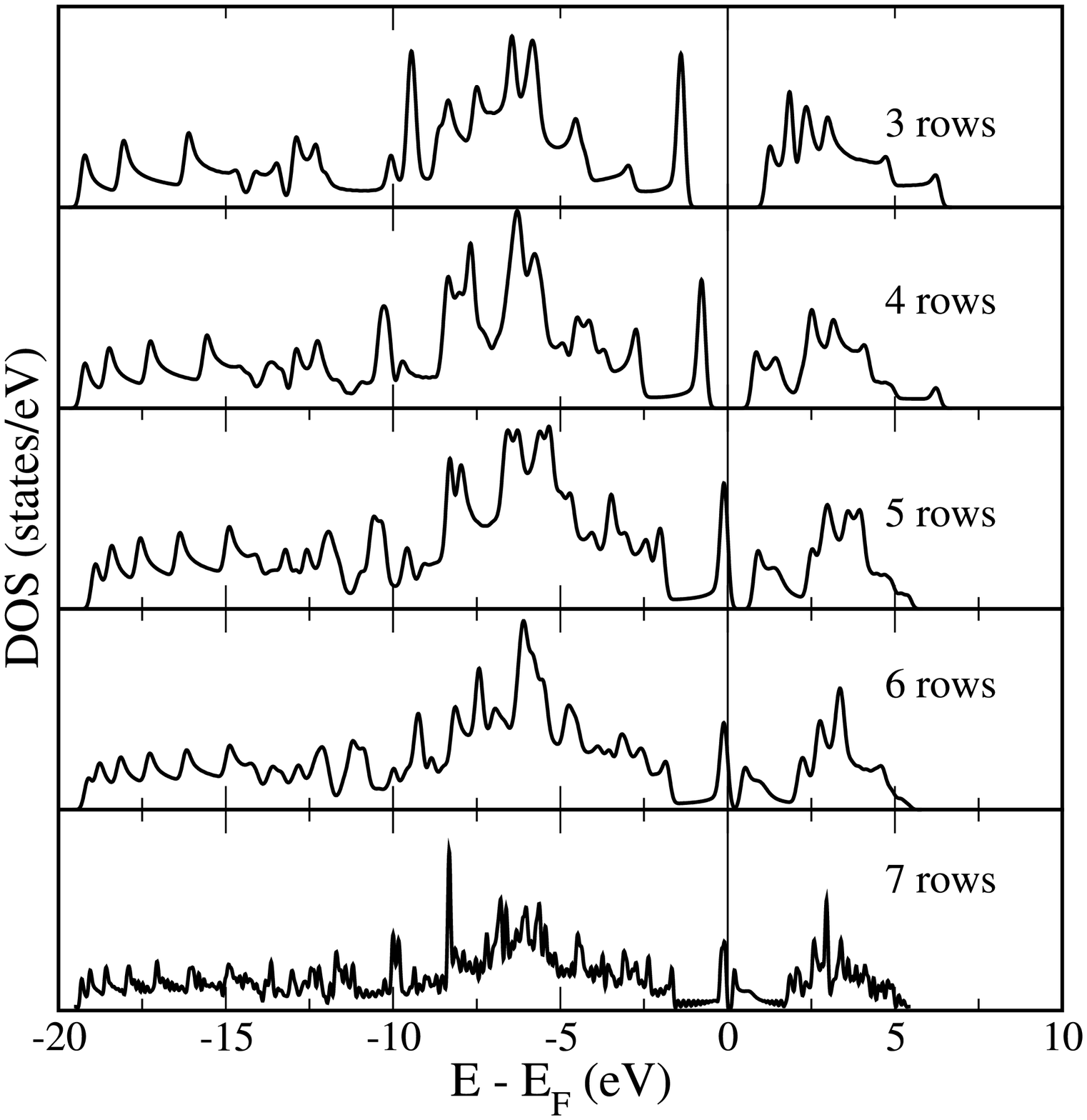}
\includegraphics[width=80mm,height=65mm]{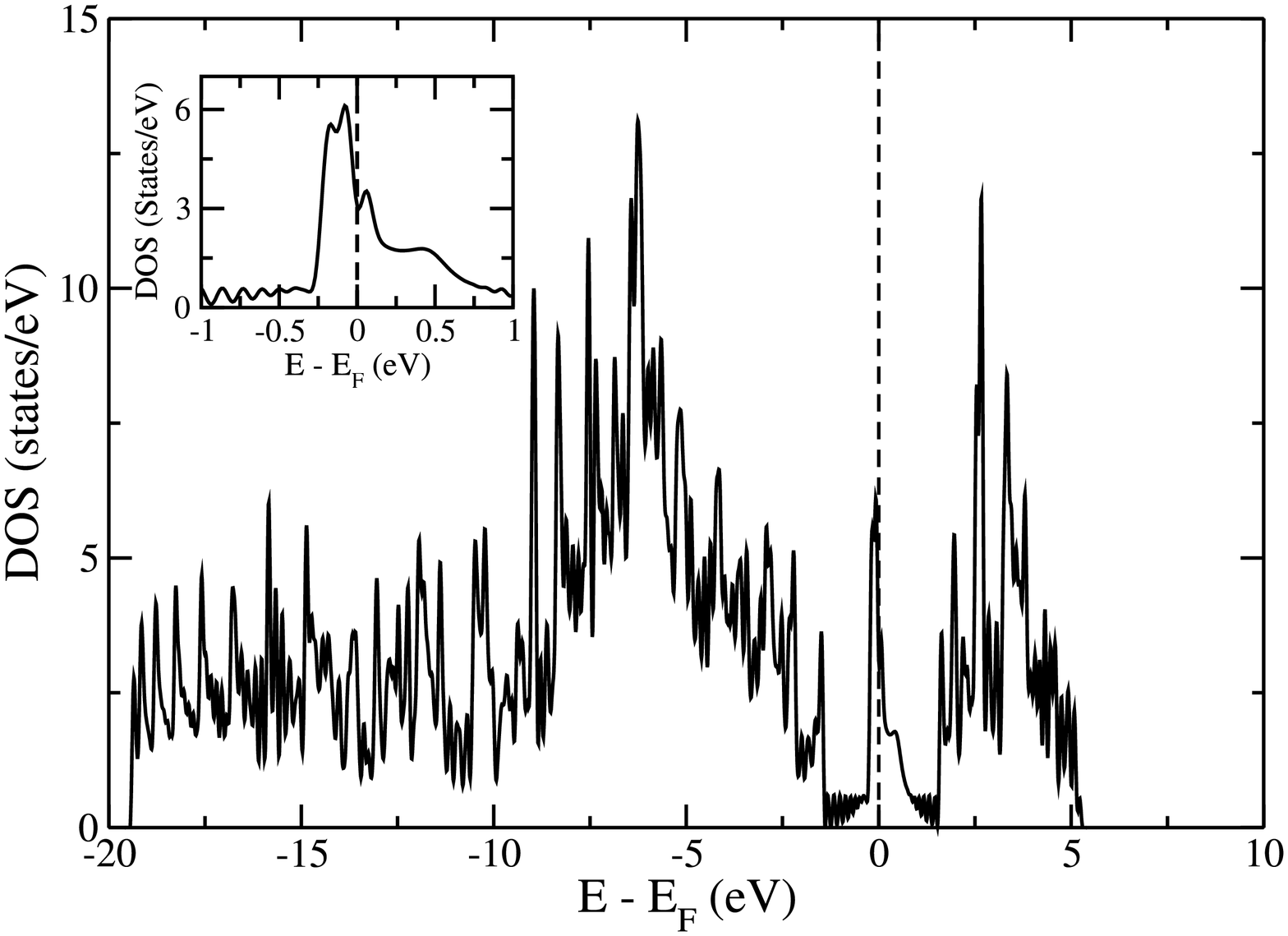}
\caption{(left) Calculated total DOS's of 2H edge terminated
nanoribbons of various widths. (right) Total DOS of an 8 row ZGNR
for a non-spin polarized calculation with (inset) expanded view of
the DOS near the Fermi level.} \label{fig4}
\end{center}
\end{figure}

Due to the formation of an sp$^{3}$ like structure at the edge and
hence the appearance of an energy gap in the electronic spectrum,
the 2H terminated ZGNR is non-magnetic. The density of states
(DOS) for different widths of ZGNR with 2H edge termination are
shown in Fig.\ref{fig4}.  One can clearly observe band gaps for 3
and 4-rows ZGNRs with values 2.1 eV and 2.08 eV, respectively. The
band gaps decrease with an increase in the ribbon width up to 7
rows ZGNR. A semiconductor to metal transition happens for an 8
row ZGNR. We observe that the Fermi level cuts through a peak in
the non-magnetic DOS (right-hand side of Fig.~\ref{fig4}). The high value of
the DOS at the Fermi level leads to an instability and hence a
spin-polarized solution leads to a lower energy state. We have
performed both non-spin polarized and spin-polarized calculations
for system sizes ranging from 8 to 20 rows ZGNR. The results are
presented in Table~\ref{tab1} where it is seen clearly that the
ribbons have a magnetic ground state, which is also metallic. This
re-entrant magnetism, which occurs despite the presence of 2H
termination is an interesting observation which we will analyze in
detail below.
\begin{table}
\caption{Difference in total energies between the non-magnetic and
ferromagnetic states ($\Delta E$) and the corresponding magnetic
moments in the ferromagnetic state. The energy differences and total magnetic moments
are quoted for the unit cell whereas the edge moment is for one C atom at the edge.}
\begin{tabular}{|c|c|c|c|}
\hline
width (rows) & $\Delta E$ (meV) & Total moment ($\mu_{B}$) & Edge moment ($\mu_{B}$) \\
\hline
\hline
8 & 1.63 & 1.04 & 0.34 \\
\hline
10 & 5.3 & 1.23 & 0.38 \\
\hline
15 & 4.9 & 1.29 & 0.39 \\
\hline
20 & 3.81 & 1.3 & 0.39 \\
\hline
\end{tabular}
\label{tab1}
\end{table}

A detailed analysis of the magnetic structure shows that the
second C atom from the edge carries most of the magnetic moment.
This second C atom from the edge develops in a non-magnetic
calculation a mid-gap state, due to the strong perturbation to the
potential its nearest neighboring C atom at the edge experiences.
The strong perturbation of the edge C atom is an effect of the
drastic difference in bonding and electronic structure of an
sp$^3$ bonded atom, compared to an sp$^2$ bonded one. For this
reason, the situation of the second C atom from the edge is quite
similar to the edge atoms of 1H terminated ZGNR. The resulting
sharp peak from the midgap state then drives the magnetic solution
of these atoms. There is one important difference between the
magnetism of the 1H terminated ZGNR and the 2H terminated ZGNR, in
that geometrical distortions of the network of sp$^3$ bonded atoms
are unlikely, and hence one does not expect Peierls-like
structural distortions, which modify the electronic structure so
that the sharp peak of the DOS at the Fermi level disappears.
\begin{figure}[h]
\includegraphics[width=80mm,height=60mm]{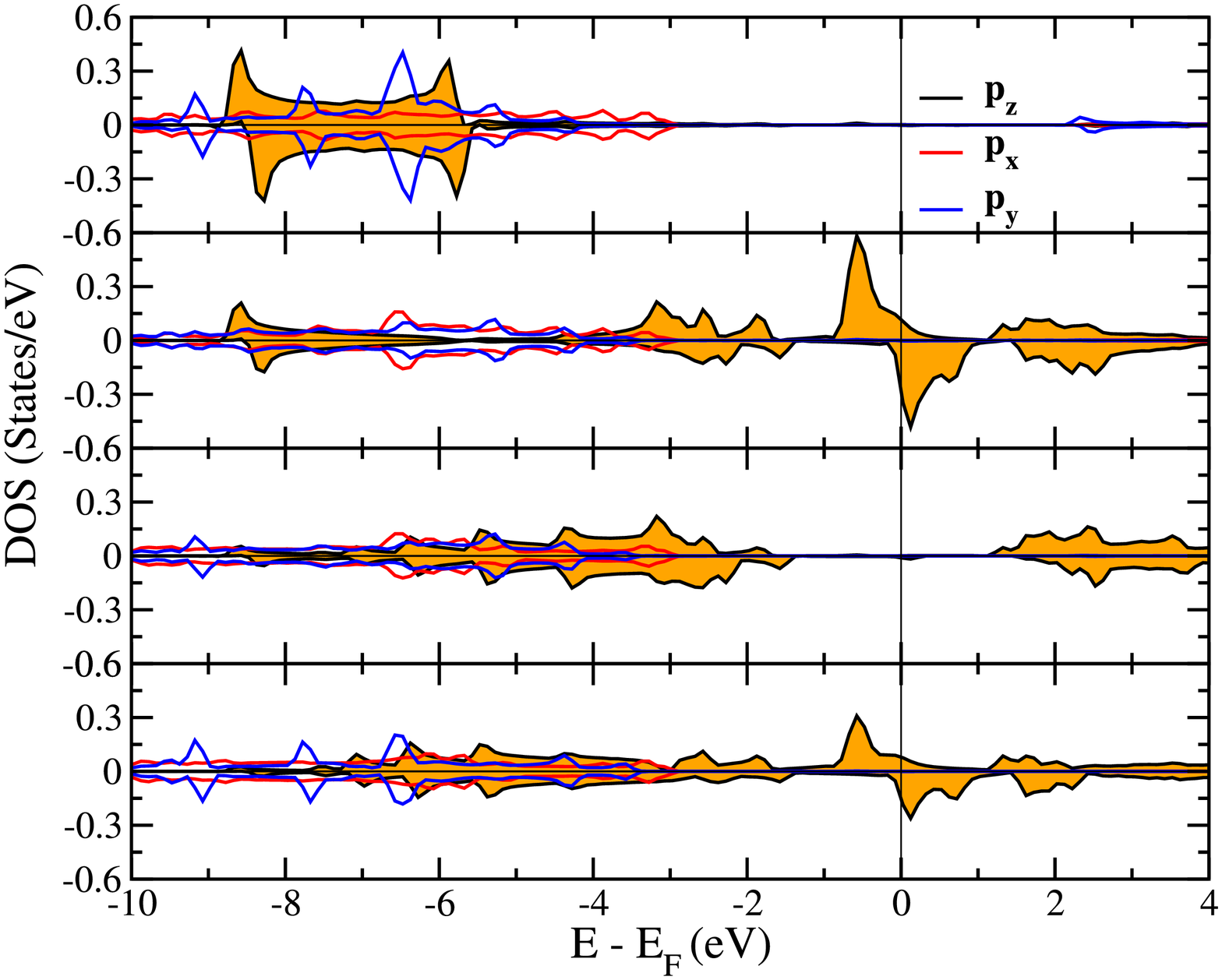}
\includegraphics[width=80mm,height=60mm]{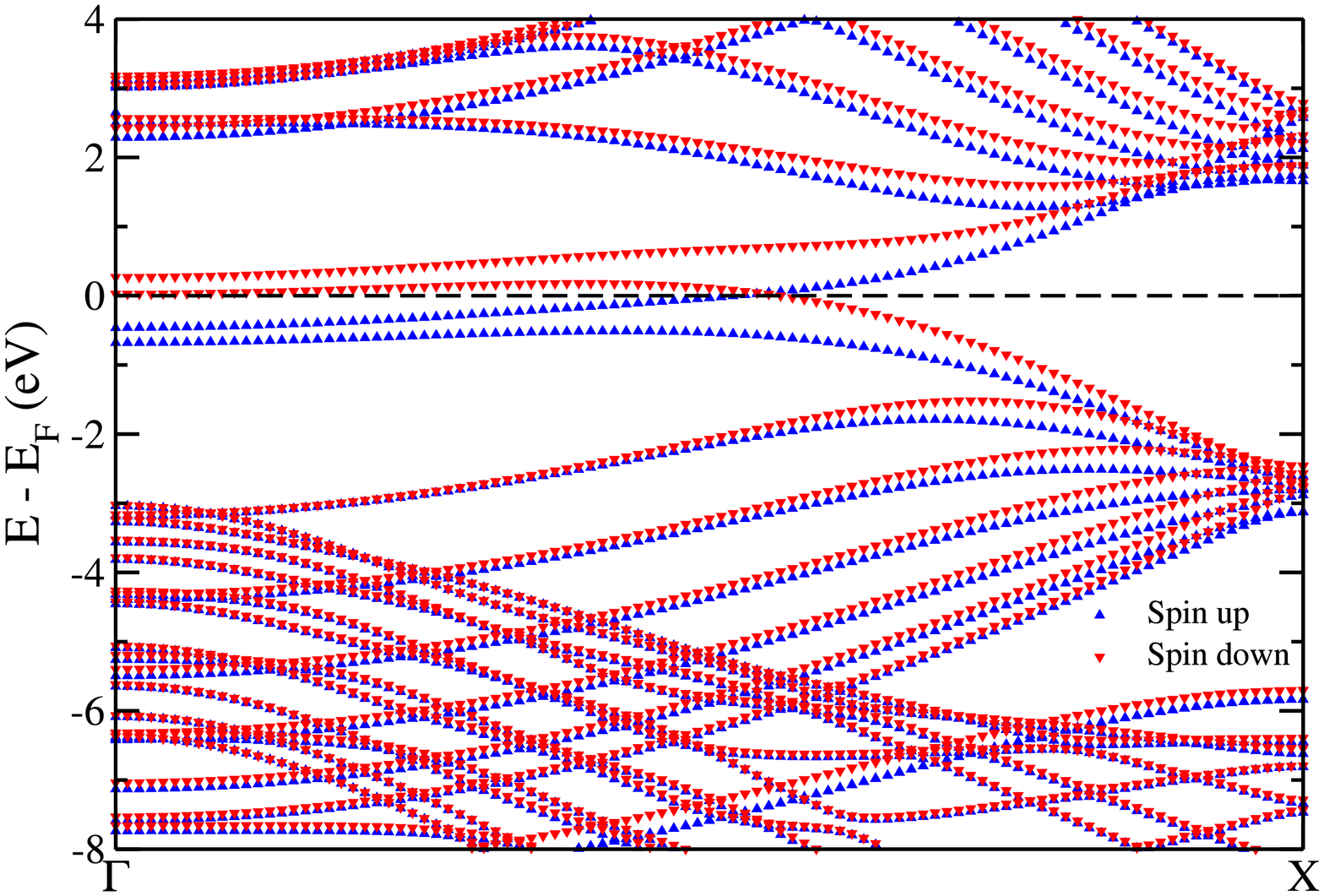}
\caption{(Color online) (left) Calculated DOS for 2H edge
terminated 8-rows nanoribbon. (Right) Spin-polarized band
structure for 2H edge terminated 8-rows zigzag nanoribbon. Both
spin-up and spin-down bands are shown.} \label{fig5}
\end{figure}

Fig.\ref{fig5} depicts the projected DOSs of the atoms starting
from the edge and going inwards for a 8-rows ZGNR. The edge atom
experiences a gap in the DOS due to sp$^{3}$ bonding and in
contrast, the DOS of the second carbon atom from the edge with a
p$_{z}$ orbital character has a sharp peak at the Fermi level
giving rise to a spin-polarized ground state due to a Stoner
instability. This alternative occurrence of a gap and a
spin-polarized state for the neighboring carbon atoms persists for
a certain distance from the edge, albeit with a decreasing
magnitude as the center of the ribbon is approached. The two
different behaviors of carbon atoms situated in different
sublattices can be described by the analysis of mid-gap states
through a tight-binding model, which we will present later on in
the paper. An inspection of the band structure (Fig.\ref{fig5}
(right)) reveals a crossing of spin-up and spin-down bands at the
Fermi level as seen in the case for a 1H terminated ZGNR.

We have also calculated the electronic structures of the 2H terminated
nanoribbons with antiferromagnetic (AFM) spins across the ribbon. The AFM state is lower in energy than the ferromagnetic state as observed for 1H terminated ZGNRs. As a result, a gap opens up at the Fermi level. As we are mostly interested in the onset of local magnetism in the wide ribbons, since long range ordered moments in low dimensional systems is hindered by the Mermin-wagner theorem, we will not discuss further the AFM state but the FM one only.
\begin{figure}[h]
\includegraphics[width=100mm,height=80mm]{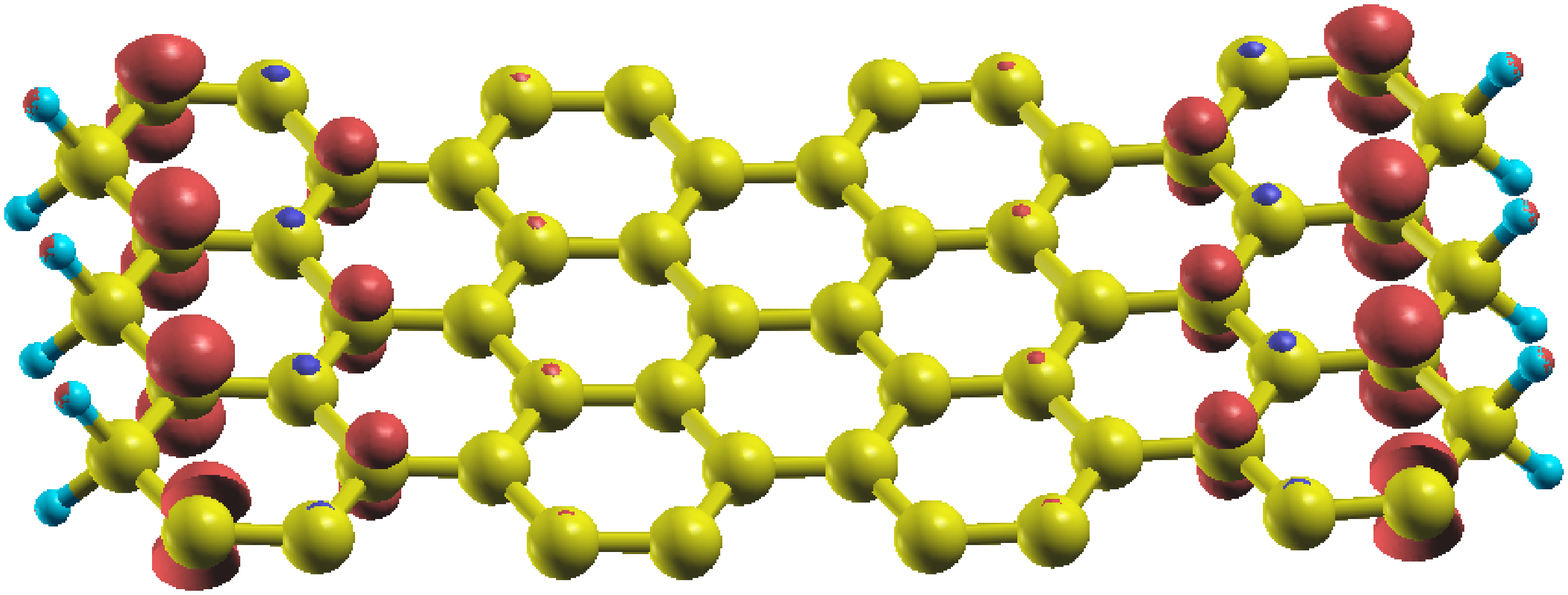}
\caption{(Color online) Isosurface plot of magnetization density of a 8-rows 2H terminated ZGNR.}
\label{fig6}
\end{figure}

In Fig.\ref{fig6}, we show the magnetization density of an 8-rows
ZGNR with 2H termination. In accordance to Fig.\ref{fig5}, we
observe the existence of a magnetization density which starts at
the nearest C neighbor to the sp$^3$ bonded C-edge atom. We also
note from the figure that only alternative carbon atoms close to
the edge of the nanoribbon have a magnetic moment and a non-zero
magnetization density.

\section{Model calculations}

\begin{figure}[h]
\includegraphics[width=80mm,height=65mm]{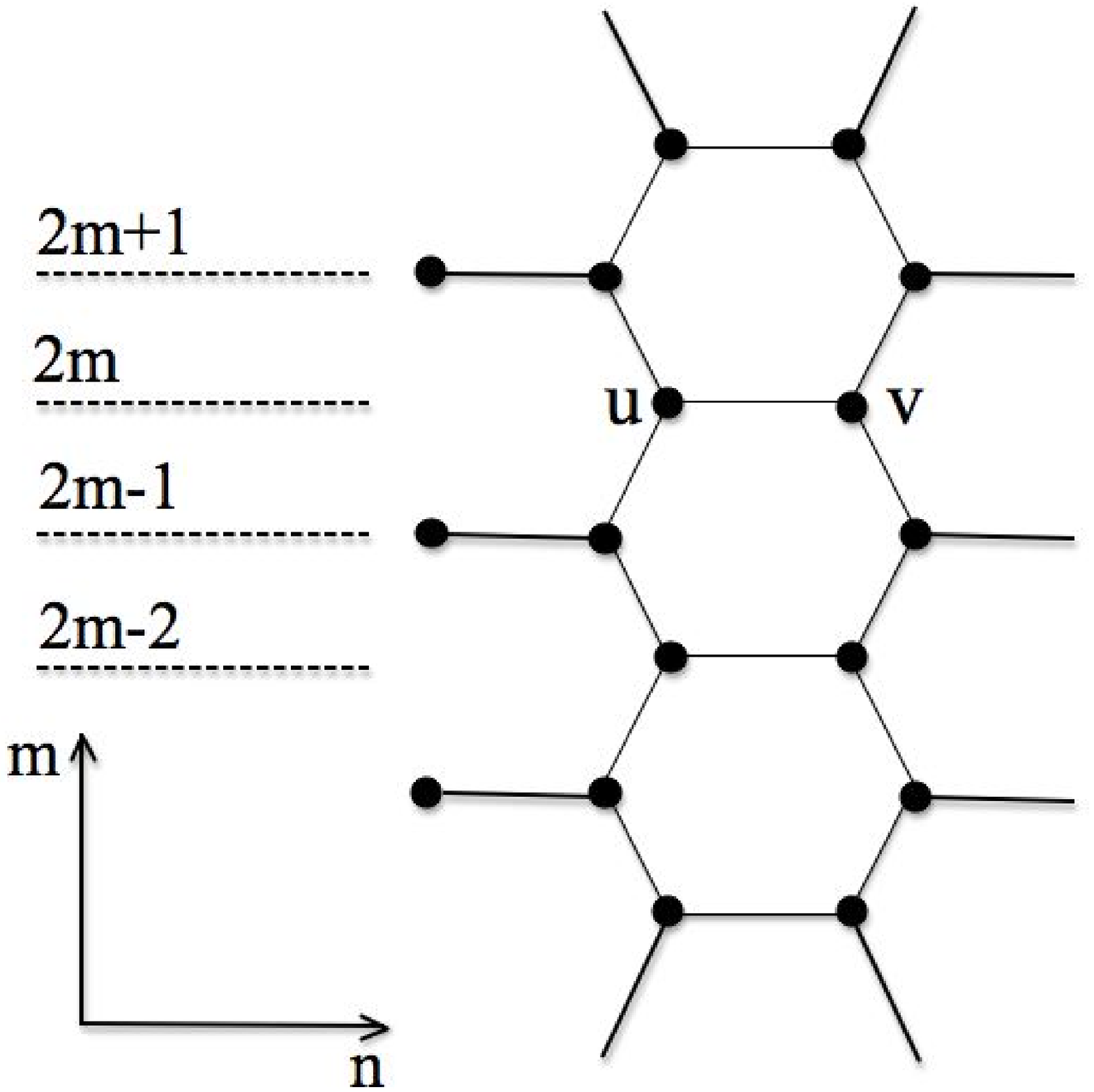}
\includegraphics[width=80mm,height=65mm]{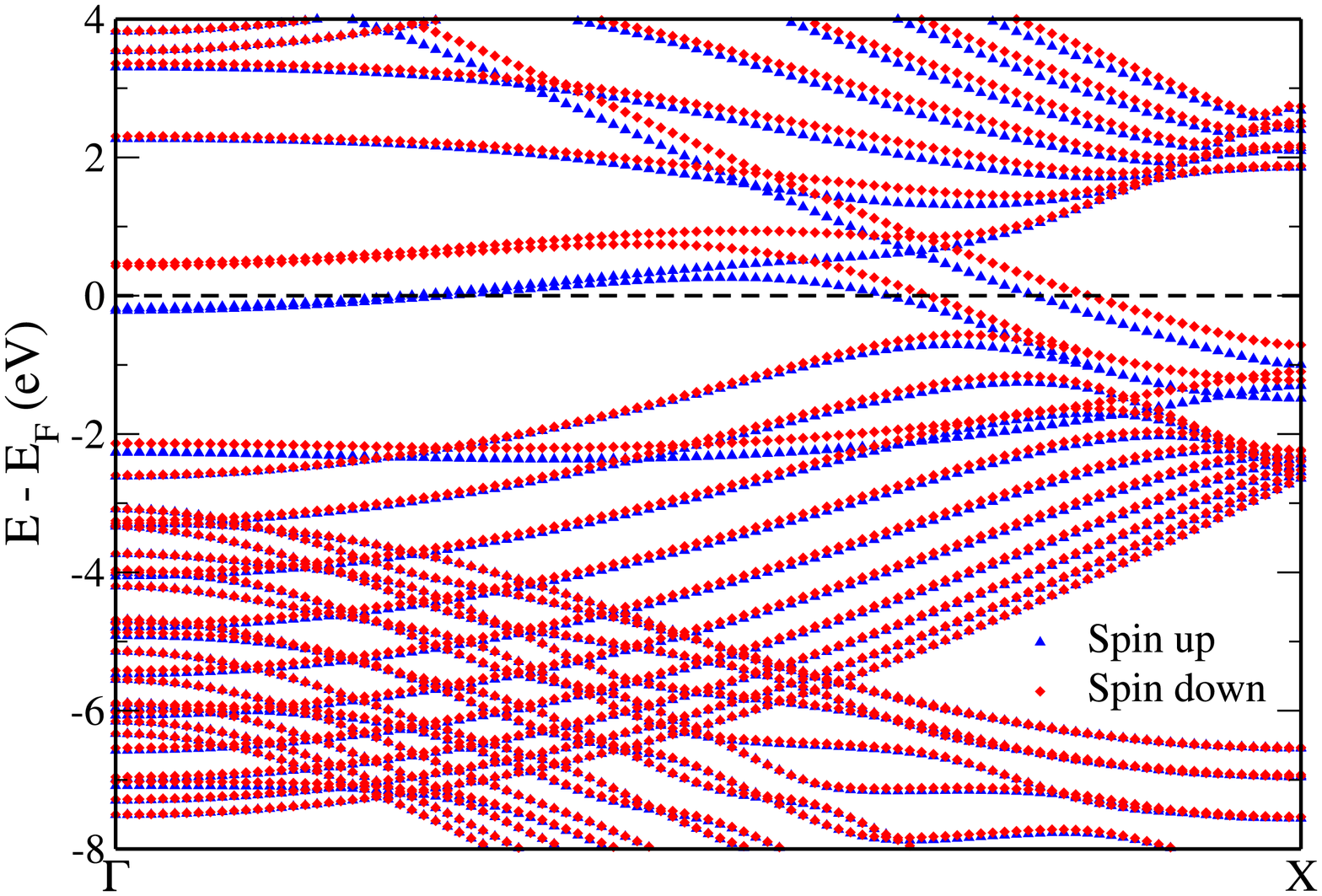}
\caption{(Color online) (left) A schematic diagram for the model
theory. (right) Spin-polarized band structure for the
carbon-terminated structure shown at the left panel. Both spin-up
and spin-down bands are shown.} \label{fig7}
\end{figure}
The first principles results presented here seem to be, at a first
look, counterintuitive. Indeed, two hydrogen atoms attached to
carbon at the zigzag edge take away all unpaired electrons, thus,
naively speaking, there is no reason to expect that the system
should become magnetic. In order to analyze the first-principles
calculations of the previous section, we have carried out analytic
calculations with a tight-binding model. A schematic diagram of
the model is shown in Fig.\ref{fig7}. In this model we introduce a
simple tight-binding Hamiltonian of the p$_z$ orbitals for
semi-infinite graphene, presented in Fig.\ref{fig7}. We assume
that the main effect of the double hydrogenation is a local gap
opening which can be approximately taken into account within the
model where the hydrogenated carbon atom is not available for the
itinerant $\pi$-bonded electrons. If we would remove all these
atoms, we would obtain a carbon-terminated structure shown in
Fig.\ref{fig7} (originally, the hydrogenated atoms were in even
rows). The extreme left carbon atoms in the model are in reality
connected with the next left row of double hydrogenated carbon
atoms, but, due to local gap opening at these atoms, within the
gap the double hydrogenated atoms are unavailable for electrons.
Our model assumption (just to interpret the ab initio results
presented above) is that the double hydrogenated carbon atoms are
unavailable for electrons at {\it any} energies and thus can be
simply removed.

In the tight-binding model with the nearest-neighbor hopping only
(the strength of the hopping parameter is put equal to 1 in our
model) the two-component wave function $(u,v)$ and the
single-electron energy $E$ can be found from the set of equations
\begin{eqnarray}
Eu\left( 2m,n\right)  &=&v\left( 2m,n\right) +v\left(
2m+1,n-1\right)
+v\left( 2m-1,n-1\right)   \nonumber \\
Ev\left( 2m,n\right)  &=&u\left( 2m,n\right) +u\left(
2m+1,n\right) +v\left(
2m-1,n\right)   \nonumber \\
Eu\left( 2m+1,n\right)  &=&v\left( 2m+1,n\right) +v\left(
2m+2,n\right)
+v\left( 2m,n\right)   \nonumber \\
Ev\left( 2m+1,n\right)  &=&u\left( 2m+1,n\right) +u\left(
2m+1,n+1\right)
+u\left( 2m-1,n+1\right)  \label{w1} \\
&&  \nonumber
\end{eqnarray}
where $n=1,2,...$ labels atoms in the direction perpendicular to
the edge. For the terminating atoms ($n=0$) one has, instead,
\begin{eqnarray}
Eu\left( 2m+1,0\right)  &=&v\left( 2m+1,0\right)   \nonumber \\
Ev\left( 2m+1,0\right)  &=&u\left( 2m+1,0\right) +u\left(
2m+2,1\right) +u\left( 2m,1\right)   \label{w2}
\end{eqnarray}
Due to translational invariance in the $m$-direction one can try the
solutions as
\begin{eqnarray}
u\left( 2m+1,n\right) =u_1\left( n\right) e^{i\xi (2m+1)}, \nonumber \\
v\left(2m+1,n\right) =e^{i\xi (2m+1)}, \nonumber \\
u\left( 2m,n\right) =u_2\left(n\right) e^{2i\xi m}, \nonumber \\
v\left( 2m,n\right) =v_2\left( n\right)e^{2i\xi m}  \nonumber \\
\label{w3}
\end{eqnarray}
One can see immediately that Eqs. 1-3 have a solution
with $E=0$, $v_1=v_2=0$ and
\begin{eqnarray}
u_1\left( n\right)  &=&\frac 1{\left( 2\cos \xi \right)
^{2n}}u_1\left(
0\right) ,  \nonumber  \label{w4} \\
u_2\left( n\right)  &=&-\left( 2\cos \xi \right) u_1\left(
n\right) . \label{w4}
\end{eqnarray}
which corresponds to the surface states at $2\left| \cos \xi
\right| >1,$ otherwise, we have nonphysical solutions growing to
the bulk. These surface states belong to only one sublattice ($u$)
and give contribution to the density of states $\frac 23\delta
\left( E\right) $ per two lines of carbon atoms (the factor 2/3 is
just a fraction of allowed values of $\xi $). When taking into
account exchange interactions between electrons, this
delta-functional peak immediately leads to the Stoner instability
and, thus, to ferromagnetism.

Eqs.(1)-(3) can be solved analytically for finite $E$ as well, but
the solutions are rather cumbersome. Instead of presenting them
here, we show in Fig.\ref{fig7} the results of first-principles
calculations for the carbon-only structure shown in the left panel
of Fig.\ref{fig7}. Indeed, we have midgap states in
non-spin-polarized case and ferromagnetism when system is allowed
to have spin polarization. Our calculated total energies reveal
that the spin-polarized solution has a lower energy compared to a
non-spin polarized one by 34.51 meV. The spin-polarized band
structure shown in Fig.\ref{fig5} has 2 spin-up and 2 spin down
bands crossing the Fermi level in contrast to the crossing of 1
spin-up and 1 spin-down bands. The analysis of the local magnetic
moments show that the edge C atom (one replacing H) has 0.3
$\mu_{B}$ and the moments decrease fast for the same sublattice C
atom. An opposite and much weaker spin-polarization is observed
for the C atom connected to the edge C atom and the same
sublattice-behavior is observed in this case too. For thick enough
ribbons, these surface states leads to the ferromagnetism which
explains our first principles results. However, if the ribbon is
very thin, the electron tunneling from one edge of the ribbon to
another one is strong enough to split the midgap state which
suppresses the spin polarization.

\section{Conclusion}
We have studied graphene nanoribbons with zigzag edges terminated
with different hydrogen concentrations to observe how the
modification along with quantum confinement and edge effects bring
changes to the electronic and magnetic properties. ZGNR with one H
atom termination per edge C atom is magnetic and an
antiferromagnetic coupling between edges stabilizes the structure
resulting a band gap which can be controlled by varying the width
of nanoribbon. Interesting features appear when the ZGNR is
terminated by 2 H atoms leading for narrow widths of the ZGNR to a
non-magnetic semiconducting system with an energy gap that
decreases with the increase of width. For sufficiently large
widths of the ZGNR, the 2 H terminated system becomes a metallic
magnet if antiferromagnetism across the ZGNR can be ordered ferromagnetically
 by an applied magnetic field. The strong sensitivity of the metallic character with respect to 
 an applied field is an interesting fact that could be relevant for sensor application 
 and spintronics. The electronic structure and magnetic state are quite
complex for this type of termination, with the sp$^3$ bonded edge
atoms being non-magnetic with a gap in the electronic structure,
whereas the nearest neighboring atoms are metallic and magnetic.
As one proceeds further in to the center of the ZGNR this trend
holds, with every row being non-magnetic and the alternating ones
having a magnetic moment. The results from ab initio theory have
been supported by a tight-binding model, which demonstrates the
presence of midgap states as the driving force for the magnetic
state. We have also evaluated the phase stability of several
thicknesses of ZGNR, and demonstrate that sp$^3$ bonded edge
atoms, with 2 H atoms at the edge, should be stable at
temperatures and pressures which are reachable in a laboratory
environment.

\section{Acknowledgement}
We gratefully acknowledge financial support from the Swedish
Research Council (VR), Swedish Foundation for Strategic Research
(SSF), G\"oran Gustafssson Foundation, Carl Tryggers Foundation, STINT, the EU-India FP-7
collaboration under MONAMI, and a KOF grant from Uppsala
University. O.E. is grateful to the ERC for support. We also
acknowledge Swedish National Infrastructure for Computing (SNIC)
for the allocation of time in high performance supercomputers. MIK
acknowledges a support from Stichting voor Fundamenteel Onderzoek
der Materie (FOM), the Netherlands.



\end{document}